\documentclass[iop]{emulateapj}

\usepackage{url}
\usepackage[utf8]{inputenc}
\usepackage{fancyref}
\usepackage{hyperref}
\hypersetup{colorlinks,breaklinks,
            urlcolor=[rgb]{0.2,0.2,0.2},  % URLs
            linkcolor=[rgb]{0.843,0.098,0.110},  % Links 
            citecolor=[rgb]{0,0.4,1.0}}  % References  
\usepackage{amssymb,amsmath}
\usepackage{enumitem}
\usepackage{natbib}
\usepackage{epstopdf}
\usepackage{rotating}
\usepackage{color}
\usepackage{pifont}
\usepackage{float}
\restylefloat{table}
\usepackage{graphicx}
\usepackage{pifont}
\usepackage{lipsum}
\DeclareMathOperator\sinc{sinc}

\definecolor{one}{rgb}{0.10588235,0.61960784,0.46666667}
\definecolor{two}{rgb}{0.85098039,0.37254902,0.00784314}
\definecolor{three}{rgb}{0.45882353,0.43921569,0.70196078}
\definecolor{four}{rgb}{0.90588235,0.16078431,0.54117647}
\definecolor{five}{rgb}{0.4,0.65098039,0.11764706}
\definecolor{six}{rgb}{0.90196078,0.67058824,0.00784314}

\newcommand{\teff}{${T}_{\mathrm{eff}}$}
\newcommand{\logg}{$\log{g}$}
\newcommand{\msun}{$M_{\odot}$}

\newcommand{\Kep}{\emph{Kepler}}
\newcommand{\Ktwo}{\emph{K2}}

\slugcomment{Accepted to ApJ}

\shorttitle{Super-Nyquist ZZ Cetis in \Ktwo}
\shortauthors{Bell et al.}

\begin{document}

\title{Destroying Aliases from the Ground and Space: \\Super-Nyquist ZZ Cetis in \Ktwo\ Long Cadence Data}
\author{Keaton~J.~Bell\altaffilmark{1,2}, J.~J.~Hermes\altaffilmark{3,4}, Z.~Vanderbosch\altaffilmark{1}, M.~H.~Montgomery\altaffilmark{1}, D.~E.~Winget\altaffilmark{1}, E.~Dennihy\altaffilmark{3}, J.~T.~Fuchs\altaffilmark{5,3}, and P.-E.~Tremblay\altaffilmark{6}}
\altaffiltext{1}{Department of Astronomy and McDonald Observatory, University of Texas at Austin, Austin, TX\,-\,78712, USA}
\altaffiltext{2}{Present Address: Max-Planck-Institut f{\"u}r Sonnensystemforschung, G{\"o}ttingen\,-\,37077, Germany}
\altaffiltext{3}{Department of Physics and Astronomy, University of North Carolina, Chapel Hill, NC\,-\,27599, USA}
\altaffiltext{4}{Hubble Fellow}
\altaffiltext{5}{Department of Physics, Texas Lutheran University, Seguin, TX\,-\,78155, USA}
\altaffiltext{6}{Department of Physics, University of Warwick, Coventry\,-\,CV4 7AL, UK}

\email{bell@mps.mpg.de}
\setcounter{footnote}{0}

\begin{abstract}

With typical periods of order 10 minutes, the pulsation signatures of ZZ Ceti variables (pulsating hydrogen-atmosphere white dwarf stars) are severely undersampled by long-cadence (29.42 minutes per exposure) \Ktwo\ observations. Nyquist aliasing renders the intrinsic frequencies ambiguous, stifling precision asteroseismology. We report the discovery of two new ZZ Cetis in long-cadence \Ktwo\ data: EPIC\,210377280 and EPIC\,220274129.  Guided by 3--4 nights of follow-up, high-speed ($\le30$\,s) photometry from McDonald Observatory, we recover accurate pulsation frequencies for \Ktwo\ signals that reflected 4--5 times off the Nyquist with the full precision of over 70 days of monitoring ($\sim$0.01\,$\mu$Hz).  In turn, the \Ktwo\ observations enable us to select the correct peaks from the alias structure of the ground-based signals caused by gaps in the observations. We identify at least seven independent pulsation modes in the light curves of each of these stars.  For EPIC\,220274129, we detect three complete sets of rotationally split $\ell=1$ (dipole mode) triplets, which we use to asteroseismically infer the stellar rotation period of $12.7\pm1.3$\,hr. We also detect two sub-Nyquist \Ktwo\ signals that are likely combination (difference) frequencies.  We attribute our inability to match some of the \Ktwo\ signals to the ground-based data to changes in pulsation amplitudes between epochs of observation. Model fits to SOAR spectroscopy place both EPIC\,210377280 and EPIC\,220274129 near the middle of the ZZ Ceti instability strip, with \teff\ $=11590\pm200$\,K and $11810\pm210$\,K, and masses $0.57\pm0.03$\,\msun\ and $0.62\pm0.03$\,\msun , respectively.

\end{abstract}


\section{Introduction}

Stellar pulsations are extremely sensitive to the detailed interior structures of stars, and Fourier analysis of photometric light curves can reveal the eigenfrequencies of these physical systems.  This is the only method by which we directly constrain stellar interiors.  However, the measurement of accurate pulsation frequencies for precision asteroseismology requires extended photometric monitoring with few gaps and short exposure times.  For data that do not meet these criteria, frequency determination is hindered by the aliasing of pulsation signals.

The atmospheres of the majority of white dwarf stars are dominated by hydrogen (DA white dwarfs).  As these stars cool, partial ionization of hydrogen eventually causes an outer convection zone to develop.  The modulation of the flux by convection can drive global nonradial oscillations \citep{Brickhill1991} that manifest as photometric variability of the star.  Near the mean mass of 0.64\,\msun\ \citep{Tremblay2013}, DA white dwarfs pulsate as ZZ Ceti variables (a.k.a.,~DAVs) in the range $12{,}500 \gtrsim {T}_{\mathrm{eff}} \gtrsim 10{,}800$ K \citep{Tremblay2013}, exhibiting pulsations with periods of 3--20 minutes.

%\LongTables
\tabcolsep=0.2cm
\begin{deluxetable*}{l c c c c c c c c}[t]
\tablecolumns{10}
\tablecaption{\Ktwo\ Target Observing Summary\label{tab:objs}}
\tablehead{
\colhead{\Ktwo\ ID}  & \colhead{Campaign} & \colhead{R.A.} & \colhead{Dec.} & \colhead{$T_0$} & \colhead{Start Date} & \colhead{Duration} & \colhead{Good Obs.} & \colhead{K$_p$} \\  
\colhead{(EPIC)} & \colhead{}  & \colhead{(h:m:s)} & \colhead{(d:m:s)} & \colhead{(BJD)} & \colhead{(UTC)} & \colhead{(d)} & \colhead{(\#)} & \colhead{(mag)}}
\startdata
210377280 & 4 & 04:04:24.924 & +12:55:43.38 & 2457061.82 & 08 Feb 2015 & 70.82 & 3249 & 18.5 \\
220274129 & 8 & 01:05:28.745 & +02:05:01.14 & 2457392.09 & 04 Jan 2016 & 78.66 & 3475 & 16.8
\enddata
\end{deluxetable*}

White dwarf asteroseismology has flourished in the era of near-continuous time series photometry from \Kep. The spacecraft saves data for pre-selected targets of interest in two observing modes: long cadence, with continuous exposures of roughly 30 minutes; and short cadence, with one-minute exposures for a limited number of targets \citep{Howell2014}. Short-cadence observations are required to sufficiently oversample typical white dwarf pulsation periods for straightforward frequency measurements.  Both in its original mission and while observing new fields along the ecliptic as \Ktwo , \Kep\ has targeted known and candidate pulsating white dwarf stars at short cadence, collecting the most extensive coverage of ZZ Ceti variability to date.  This has enabled the precise determination of pulsation frequencies for asteroseismic analysis \citep{Greiss2014,Hermes2014,Hermes2015a,Hermes2017,Hermes2017b,Bell2015}, as well as the discovery of a pulsation-related outburst phenomenon that operates near the cool edge of the ZZ Ceti instability strip \citep{Bell2015,Bell2016,Bell2017,Hermes2015b}.

Long-cadence \Ktwo\ observations can potentially reveal white dwarf pulsations, but the signals suffer dramatic aliasing against the Nyquist frequency, as well as amplitude reduction, since the 30-minute cadence severely undersamples the pulsations.  While the sub-Nyquist pulsation frequency aliases measured from these data are extremely inaccurate, they are of exceptionally high precision owing to the long observational baseline of \Ktwo .

High-speed photometry from individual ground-based observatories causes a different kind of aliasing: gaps in the data from daylight and weather introduce cycle-count ambiguities into the pulsation record. Selecting the correct alias among the comb of peaks that make up the observational spectral window is nontrivial.

By combining long-cadence \Ktwo\ data with single-site high-speed follow-up from McDonald Observatory, we can use the strengths of each data set to resolve the aliasing in the other. We take this approach in analyzing two new ZZ Ceti variables: EPIC\,210377280 from \Ktwo\ Campaign 4 and EPIC\,220274129 from Campaign 8. We describe the \Ktwo\ observations, as well as follow-up spectroscopy and high speed photometry, in Section~\ref{sec:obs}.  We discuss the instrumental and astrophysical effects that cause differences in the pulsation signatures between our ground- and space-based light curves in Section~\ref{sec:effects}.  We then carefully analyze these data sets together for both EPIC\,210377280 and EPIC\,220274129 in Section~\ref{sec:analysis}, identifying multiple significant pulsation frequencies in each star. We recover \Ktwo -level precision for those modes that we are able to match between data sets. We discuss the \Ktwo\ signals that are not matched in the ground-based light curves in the concluding Section~\ref{sec:dc}.

\section{Observations}
\label{sec:obs}

\subsection{Long-cadence \Ktwo\ photometry}

EPIC\,210377280 and EPIC\,220274129 were observed at long cadence in \Ktwo\ fields 4 and 8 for 70.82 and 78.66 days, respectively, as part of a search for white dwarf transits and variability with \Ktwo . Both were high-probability white dwarf candidates from the catalog of \citet{GF2015}, but they were not included in our short-cadence survey for pulsating white dwarfs since their ($u$$-$$g$, $g$$-$$r$) colors were well outside our usual selection region for typical ZZ Cetis. Both stars exhibit multiple significant periods of variability in the long-cadence \Ktwo\ photometry.

For each target, we utilize the light curves extracted and processed by the EVEREST 2.0 pipeline \citep{Luger2016,Luger2017}, discarding all points with ``quality'' flags set. We flatten the light curves further by dividing by a cubic spline fit\footnote{Using tools from {\sc SciPy}: \url{http://www.scipy.org/}}, tuning the fitting weights manually to remove residual noise at the lowest frequencies without affecting the measured amplitudes of the signals of astrophysical interest in the Fourier transform.  We note that version 2.0 of the EVEREST pipeline accounts for contamination from nearby stars in the photometric aperture, which otherwise could dilute the amplitudes of variability measured  in \Kep 's $4\arcsec$ pixels.

The details of the \Ktwo\ observations of both stars are summarized in Table~\ref{tab:objs}, including the total number of long-cadence observations in the final light curves.

\subsection{Time series photometry from the ground}

We confirmed the pulsational nature of the photometric variations of both targets with high-speed follow-up from the ground with the ProEM camera\footnote{\url{https://www.princetoninstruments.com/products/ProEM-EMCCD}} on the 2.1-m Otto Struve Telescope at McDonald Observatory. At Cassegrain focus, the ProEM camera has a 1.6$\arcmin$$\times$1.6$\arcmin$ field of view, and with 4$\times$4 binning, our effective plate scale is 0.38$\arcsec$ pixel$^{-1}$. Each object was observed on four different nights, as detailed in Table~\ref{tab:obs}. The data were acquired through a broad-band BG40 filter that transmits between 3300--6000\,\AA\ to reduce sky noise.

We use standard {\sc IRAF} tasks to dark-subtract and flatfield the images with calibration frames that we acquired each night.  We extract circular aperture photometry for the target and comparison stars in the field using {\sc ccd\_hsp}, an {\sc IRAF} script that uses tasks from {\sc phot} \citep{Kanaan2002}.  We divide the measured target counts by a weighted sum of comparison star counts, then by a low-order best-fit polynomial to correct for airmass and transparency variations using the {\sc Wqed} tools \citep{Wqed}.  {\sc Wqed} also applies a barycentric correction and accounts for the latest leap seconds to enable the reliable combination of multiple nights of data.

\begin{deluxetable}{l r c c}[t]
\tablecolumns{4}
\tablecaption{Journal of McDonald Observations\label{tab:obs}}
\tablehead{
\colhead{EPIC} & \colhead{Date} & \colhead{Exposure} & \colhead{Run Duration} \\
\colhead{} & \colhead{(UTC)} & \colhead{Time (s)}	& \colhead{(h)}}
\startdata
210377280    & 04 Feb 2016 & 15 & 4.6 \\
                        & 05 Feb 2016 & 10 & 4.5 \\
                        & 07 Feb 2016 & 20 & 1.9 \\
                        & 07 Feb 2016 & 10 & 2.3 \\
                        & 08 Feb 2016 & 15 & 4.7 \\
220274129    %& 07 Jul 2016 & 10 & 1.4  \\
                        & 28 Oct 2016 & 3 & 5.5  \\
                        & 29 Oct 2016 & 20 & 2.0  \\
                        & 29 Oct 2016 & 30 & 3.0  \\
                        & 01 Nov 2016 & 10 & 6.0
\enddata
\end{deluxetable}

\begin{figure}[t]
  \centering
  \includegraphics[height=0.99\columnwidth,trim={4cm 5.5cm 5.6cm 6.9cm},clip,angle=-90]{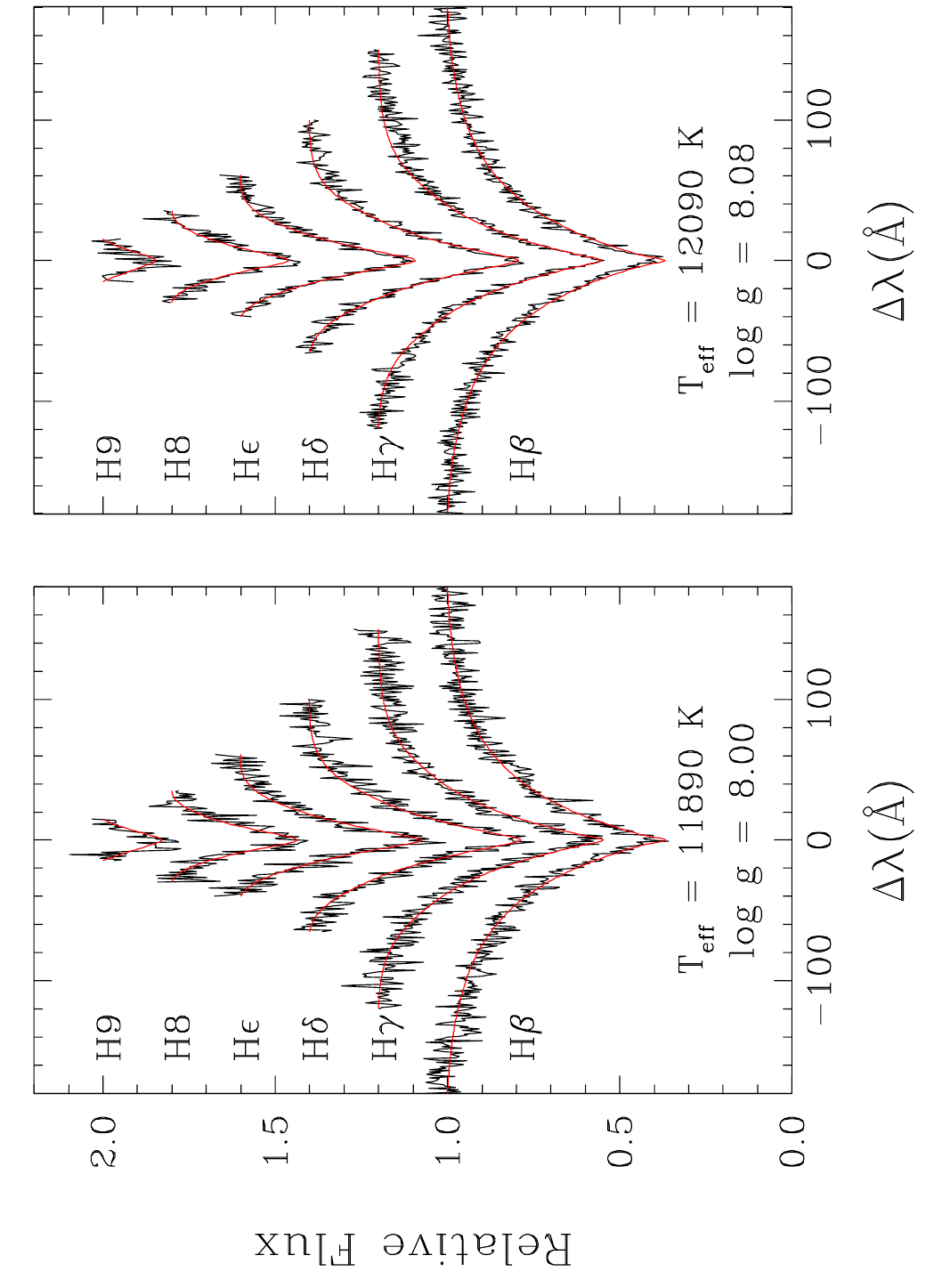}
  \caption{Balmer line profiles from the SOAR spectra of EPIC\,210377280 (left) and EPIC\,220274129 (right). The best fit 1D models \citep[described in][]{Tremblay2011} are overplotted in red, and their parameters are indicated at the bottom of each panel.}
  \label{fig:spec}
\end{figure}

\subsection{Spectroscopy}
\label{sec:spec}

\begin{deluxetable*}{l r c c c r}[b]
\tablecolumns{9}
\tablecaption{Spectroscopic Parameters \label{tab:spec}}
\tablehead{
\colhead{EPIC} & \colhead{\teff} & \colhead{\logg} & \colhead{\teff} & \colhead{\logg} & \colhead{Mass} \\
\colhead{} & \colhead{(K; 1D)} & \colhead{(cgs, 1D)}	& \colhead{(K; 3D)} & \colhead{(cgs, 3D)}	& \colhead{(\msun)} }
\startdata
210377280   & 11890(200)  &  8.00(0.06)  & 11590 & 7.94 & 0.57(0.03) \\
220274129   & 12090(210)  &  8.08(0.05) & 11810 & 8.03 & 0.62(0.03)
\enddata
\end{deluxetable*}

We acquired spectroscopic observations for both stars with the Goodman spectrograph \citep{Clemens2004} on the 4.1-m SOAR telescope on Cerro Pach\'on, Chile, using the setup described in \citet{Hermes2017b}. The EPIC 210377280 and EPIC 220274129 spectra were obtained on 02 Sept and 14 July 2016, respectively,

We fit the data to 1D atmosphere models following the methodology of \citet{Tremblay2011}, with the ML2/$\alpha = 0.8$ mixing length theory parameterization. The uncertainties include the external errors
from data reduction and flux calibration estimated by \citet{Gianninas2011}.  We also apply the 3D convection corrections of \citet{Tremblay2013}. Both sets of atmospheric parameters are provided in Table~\ref{tab:spec}.  These parameters place both stars near the middle of the ZZ Ceti instability strip \citep{Tremblay2013}. A comparison of the best-fit models to the observed Balmer line profiles is displayed in Figure~\ref{fig:spec}.

\section{Comparability of Data Sets}
\label{sec:effects}

There are a number of effects, both observational and astrophysical, that cause the signatures of pulsations in these stars to differ between the \Ktwo\ photometry and the follow-up ground-based light curves.  We must factor in these considerations as we approach the task of matching signals between these data sets.  In this section, we discuss the most significant of these effects and how we incorporate them into the analysis that follows.

\subsection{Nyquist aliasing}
\label{sec:nyq}

\begin{figure}[t]
  \centering
  \includegraphics[width=0.99\columnwidth]{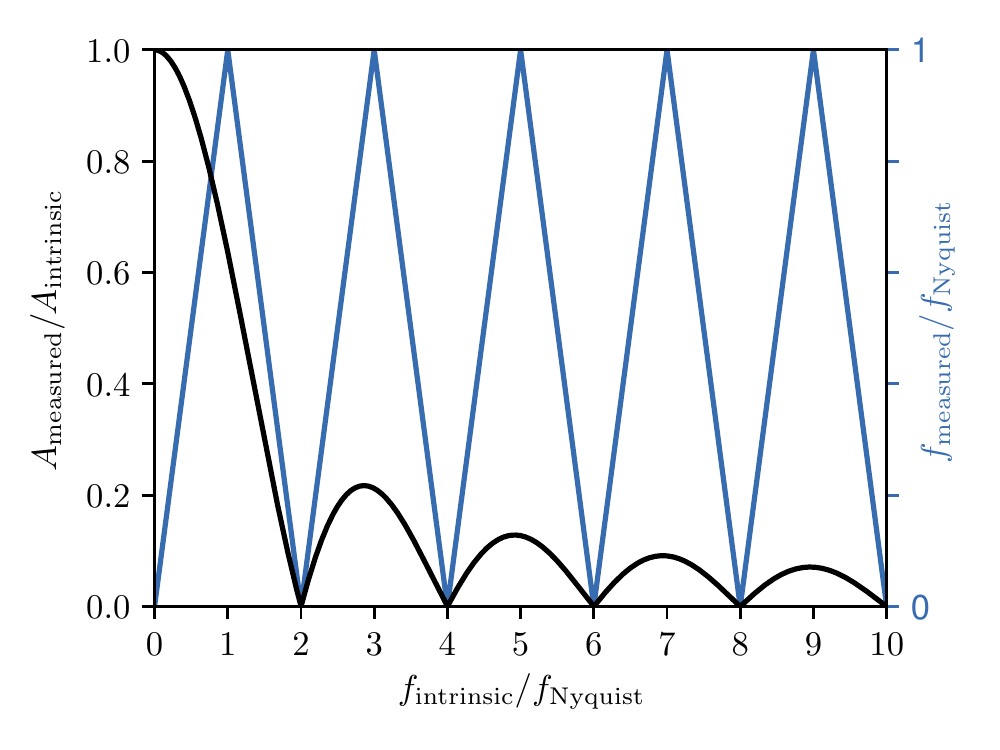}
  \caption{The effect of continuous time series sampling on the measured amplitude and frequency of a signal. Signals with intrinsic frequencies beyond the Nyquist frequency will be aliased into the sub-Nyquist regime (right axis) with significantly decreased amplitudes (left axis).}
  \label{fig:sn}
\end{figure}

The Nyquist critical frequency for evenly sampled data is $f_\mathrm{Ny} = 1/(2\Delta t)$, where $\Delta t$ is the constant spacing in time between observations.  Fourier transforms (FTs) of photometric light curves exhibit peaks at the accurate frequencies of intrinsic stellar brightness variations for signals that are bandwidth limited to below $f_\mathrm{Ny}$.  However, frequencies of stellar photometric variability greater than $f_\mathrm{Ny}$ will be aliased into the range $0 < f < f_\mathrm{Ny}$. Figure~\ref{fig:sn} demonstrates the relationship between the underlying signal frequency and the location of its $0 < f < f_\mathrm{Ny}$ alias.

The long-cadence \Ktwo\ light curves are acquired onboard the spacecraft with a regular spacing of 29.43 minutes.  These timestamps are then corrected for the changing distance between the spacecraft and the solar system barycenter, breaking the strict regularity of the time sampling. The Nyquist frequency is not so simply defined under these conditions.  \citet{Eyer1999} show that the effective Nyquist frequency over which signals are exactly aliased is $f_\mathrm{Ny} = 1/{(2p)}$, where $p$ is the greatest common factor of all time separations between pairs of observations. $p$ is the longest period that the time series could be folded on to cause all samples to be coincident in phase. \citet{Koen2006} points out that there is a practical lower limit on $p$ (upper limit on $f_\mathrm{Ny}$) set by the recorded timestamp precision.

FTs of the \Ktwo\ data are significantly affected by Nyquist-like aliasing across lower frequencies due to the near-even spacing of \Ktwo\ observations.  This pseudo-Nyquist behavior occurs at $f_\mathrm{Ny}^\ast = 1/{(2p^\ast)}$, where the time samples folded on $p^\ast$ are highly concentrated in phase.  \citet{Murphy2013} exploited the difference in aliasing behavior between this and the true Nyquist frequency to identify the intrinsic frequencies for super-Nyquist pulsators in the original \Kep\ Mission data.

Following the framework of \citet{Eyer1999}, we identify the pseudo-Nyquist frequency by direct calculation of the spectral window (the signature in the FT of a pure sinusoid sampled as the data). The frequency of the first peak beyond zero in the spectral window with amplitude near unity is twice $f_\mathrm{Ny}^\ast$. The pseudo-Nyquist frequencies of EPIC\,210377280 and EPIC\,220274129 are 283.2377 and 283.2388 $\mu$Hz, respectively. The distinction is subtle, so we refer to $f_\mathrm{Ny}^\ast$ as simply the Nyquist frequency for the remainder of this paper.

\subsection{Phase smearing}
\label{sec:ps}

Finite exposure lengths of photometric observations have the effect of boxcar smoothing the underlying signal.  The amplitudes measured from the data are therefore smaller than the intrinsic signal amplitudes, and a correction factor must be applied if accurate amplitudes are of interest.  When exposures of duration $t_{\mathrm{exp}}$ are used to record a sinusoid of period $P$, its amplitude will appear smaller by a factor of $\eta$:
\begin{equation}
\eta = A_{\mathrm{measured}}/A_{\mathrm{intrinsic}} = \sinc{(\pi t_{\mathrm{exp}}/P)}.\footnote{$\sinc{(x)}=\sin{(x)}/x$}
\end{equation}

Figure~\ref{fig:sn} demonstrates this in terms of the ratio of the signal frequency to the Nyquist frequency, assuming continuous exposures with no overhead. This expression for phase smearing can be found in various works, including \citet{Hekker2016}, where it is referred to as ``apodization.'' A derivation is available in \citet[Section 1.2.2]{Murphy2014}.

This effect has a major impact on the ZZ Ceti pulsation amplitudes measured from long-cadence \Ktwo\ data, where $t_{\mathrm{exp}} > P$.  The equation also expresses \Ktwo's lack of sensitivity to signals with periods that are near integer fractions of the exposure time.

\subsection{Passband differences}
\label{sec:pd}

Variations of the emergent flux from a star due to pulsations are wavelength dependent.  This causes the same pulsations to have different measured amplitudes in different photometric systems.  \emph{Kepler} has a broad response function spanning roughly 4300--8900\,\AA \footnote{\url{https://keplerscience.arc.nasa.gov/CalibrationResponse.shtml}}, while the McDonald observations were made through Earth's atmosphere and a BG40 filter that spans 3300--6000\,\AA .

To account for this, we calculated the expected pulsation amplitude ratio as measured through these two passbands for a {representative ${T}_{\mathrm{eff}}=11{,}700$\,K, $\log{g}=8.0$ DA white dwarf (within $1\sigma$ of the 3D-corrected spectroscopic parameters of both stars).  We utilized the grid of spectroscopic model atmospheres described in \citet{Koester2010}---with a ML2/$\alpha = 0.8$ treatment of convection---to model the emergent flux.} We consider pulsations of spherical degree $\ell =1$ and $\ell=2$, since practically all available constraints on white dwarf pulsations are consistent with these two spherical degrees \citep[e.g.,][]{Winget2008}. {We expect the intrinsic amplitudes of $\ell =1$ and $\ell=2$ pulsation modes to be larger in the McDonald data than in the \Ktwo\ observations by factors of roughly 1.126 and 1.133, respectively.  We account for this effect by applying the average scaling factor of 1.130} when comparing measurements between data sets throughout this work.

\subsection{Spectral window}
\label{sec:sw}

Gaps in data can cause confusion in the determination of pulsation frequencies by introducing an uncertainty in the number of cycles missed when observations were not being made.  Our ground-based data suffer significant diurnal aliasing since they are distributed across multiple nights.  As opposed to the intrinsic limitations on frequency precision set by the spectral resolution \citep[$\propto 1/T$, where $T$ is the total baseline of observations;][]{Montgomery1999}, aliasing introduces an additional \emph{extrinsic} error to frequency determinations, since we might select an inaccurate alias peak.  

The uncertainty introduced by these aliases is best understood by studying the spectral window: the signature of a pure sinusoid in the FT that arises solely from the time sampling. We plot the spectral windows for both stars as part of our analyses that follow.  We characterize the magnitude of the potential frequency confusion by measuring the location of the highest non-central alias in the spectral window.

Any pulsation modes included in our frequency solutions based on the ground-based data alone are flagged as potentially suffering from spectral window alias ambiguities.  We emphasize that this extrinsic error is non-Gaussian, and that any attempt at asteroseismic inference should carefully consider alternate frequency solutions for these stars that employ other viable aliases.

\subsection{Intrinsic mode variation}
\label{sec:coherence}

Despite our best efforts to account for how differences in instrumentation affect the comparability of pulsation signatures between these data sets, we face an astrophysical limitation that is more difficult to surmount: the intrinsic signatures of the stellar pulsations may have changed significantly between epochs of observation.  The challenges of comparing multi-season pulsation measurements for ZZ Ceti variables that have cooled significantly beyond the hot edge of the instability strip are well established in the literature, particularly from Whole Earth Telescope \citep{Nather1990} campaigns. GD\,154 \citep{Pfeiffer1996}, G29--38 \citep{Kleinman1998}, HL~Tau~76 \citep{Dolez2006}, and EC14012-1446 \citep{Handler2008,Provencal2012}, for example, show such dramatic amplitude changes that significant modes from one year can be completely absent the next. With nine and seven month gaps between space- and ground-based campaigns on EPIC\,210377280 and EPIC\,220274129, we have no expectation that the same eigenmodes will be excited to similar amplitudes in both data sets.

Extended, continuous records from \Kep\ and \Ktwo\ have provided new insights into mode variations \citep[e.g.,][]{Hermes2014,Bell2015,Bell2016}.  By inspecting the FTs of 27 ZZ Cetis observed by \Kep\ and \Ktwo , \citet{Hermes2017b} discovered a dichotomy of mode behavior: while some modes are notably coherent, most with periods longer than 800\,s undergo significant modulation and appear as multiple closely spaced peaks in the FT.  These bands of power are well fit by Lorentzians with half-widths-at-half-maximum of order 1\,$\mu$Hz. The amplitudes of the many individual peaks that make up these power bands are typically smaller than the instantaneous, intrinsic amplitudes of the corresponding pulsation modes.  Unlike for phase smearing and passband effects, we are unable to apply a simple corrective factor that accounts for this mode incoherence.

\section{Comparing Data Sets}
\label{sec:analysis}

The ZZ Ceti pulsation signals in long-cadence \Ktwo\ data may have suffered any integer number of Nyquist reflections.  We assess these candidate intrinsic frequencies through comparison with ground-based observations from McDonald Observatory. Pulsation frequencies that we are able to positively match between these data sets do not suffer measurement ambiguities from either the Nyquist aliasing of the \Ktwo\ data or the window function aliasing of the multi-night ground-based photometry.  We conservatively accept only unique, unambiguous matches between data sets.

To determine the signals of statistical significance in the \Ktwo\ data, we use a bootstrapping approach to calculate a threshold in the FT that corresponds to a 0.1\% false alarm probability (FAP). This is similar to the calculations made in, e.g., \citet{Greiss2014}, \citet{Bell2015,Bell2016}\footnote{We note that the application in these previous works was not strictly bootstrapping, as the resampled flux values were drawn without replacement (i.e., shuffled). The 0.1\% FAP thresholds calculated through resampling with replacement agree to within 1\% the values from resampling without replacement.}.  Signals are significant to better than 99.9\% confidence\footnote{This is a slightly conservative criterion, since this approach treats astrophysical signals as additional sources of noise.} if they exceed the 99.9 percentile of maximum peak amplitudes detected in the full FTs of 10{,}000 random bootstrap samplings of the \Ktwo\ light curves of each star.

The rule-of-thumb significance criterion for ground-based light curves is {to reject the null hypothesis (that peaks can be ascribed to pure noise) for peaks in excess of four times the mean amplitude evaluated in a local region of the FT ($4\langle A\rangle$; \citealt{Breger1993}).  These peaks would be accepted as signal with a FAP of $\approx$\,0.1\% \citep{Kusching1997}.} For these data sets, a high density of pulsation signatures convolved with a broad spectral window makes {this evaluation of the local noise level} a challenge, and we instead interpolate between $4\langle A\rangle$ values calculated in regions on each side of the frequency range of pulsational power: 0--800\,$\mu$Hz and 6000--8000$\,\mu$Hz.

All FTs examined in this work were computed with the \mbox{\sc Period04} software package \citep{Period04}.  We oversample the spectral resolution by a factor of 20 to obtain representative peak amplitudes. We average the ground-based light curves into 60\,s bins (matching the least common multiple of exposure times used on different nights) to avoid giving runs with shorter exposure times undue weight in the FT.

While our primary analysis utilizes the \Ktwo\ light curves extracted by the EVEREST pipeline, we test the robustness of our signal detections by confirming their presence in the {PDC\_SAP} light curves\footnote{\url{https://keplerscience.arc.nasa.gov/k2-data-release-notes.html}} produced by the {\Ktwo\ Pipeline} \citep{VanCleve2016} and the extractions described by \citet{VJ2014}.

\subsection{EPIC\,210377280}
\label{sec:star1}

\begin{figure}[t]
  \centering
   \includegraphics[width=.9\columnwidth]{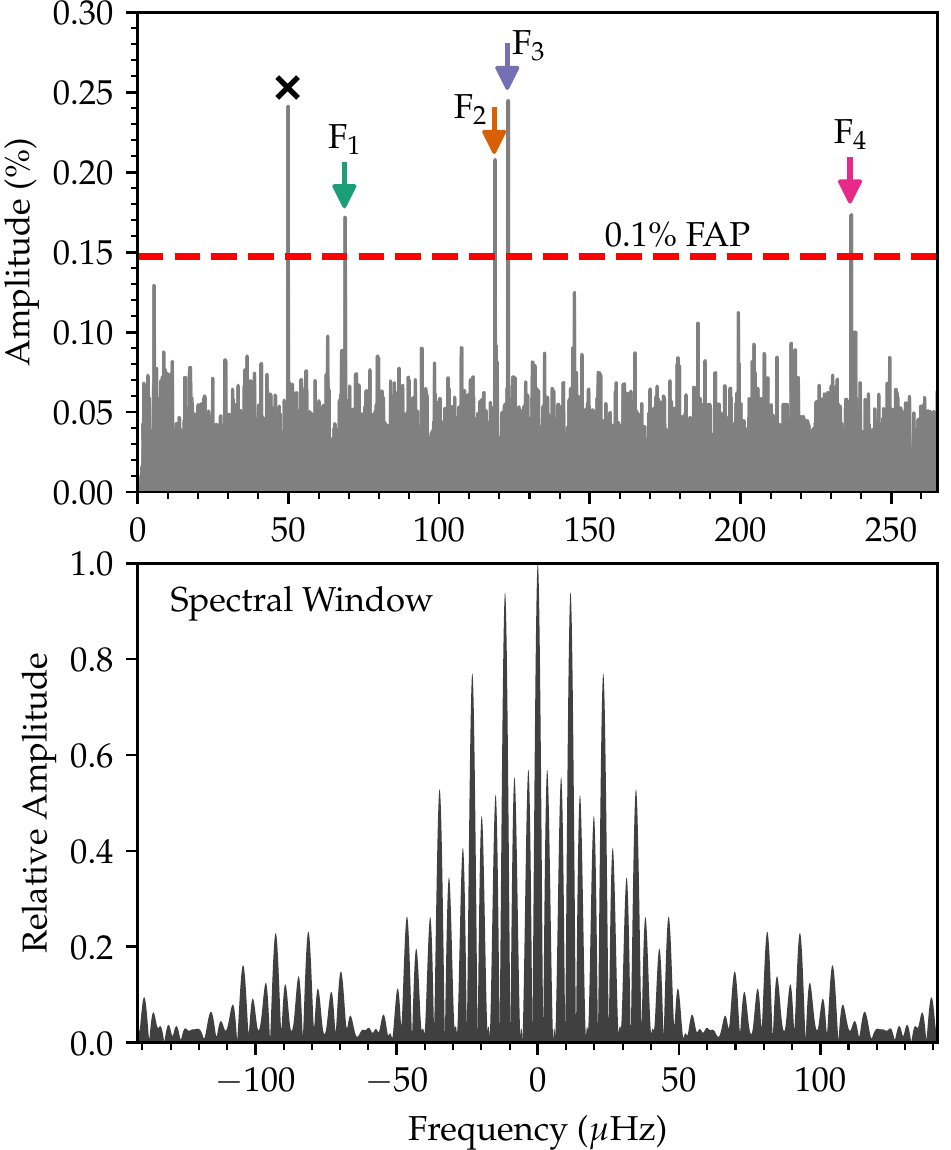}
  \caption{{\sc Top:} Fourier transform of the \Ktwo\ observations of EPIC\,210377280 out to the Nyquist frequency with a 0.1\% false alarm probability (FAP) significance threshold (see text). Significant signals are marked with colored arrows. {\sc Bottom:} The spectral window from four nights of McDonald observations of EPIC\,210377280 over a five night span in Feb 2016. The highest alias is located 11.6\,$\mu$Hz away from the central peak. The x-axes of both panels have the same scale.}
  \label{fig:threshsw1}
\end{figure}

The FT of the long-cadence \Ktwo\ light curve of EPIC 210377280 is displayed in the top panel of Figure~\ref{fig:threshsw1}.  Our bootstrap calculation yields a 0.1\% FAP significance threshold at 0.148\% amplitude, which is indicated by the dashed red line.  The peak marked with a $\times$ symbol is within 0.3\,$\mu$Hz of a typical instrumental artifact at $\approx$50\,$\mu$Hz caused by \Ktwo's thruster firings \citep{VanCleve2016}, so we exclude it from our analysis. We mark four significant signals of astrophysical interest with colored arrows. Their properties, computed from least-squares fits to the entire \Ktwo\ light curve, are listed in Table~\ref{tab:k21}.

The spectral window from four nights of ground-based observations of EPIC\,210377280 from McDonald Observatory is displayed in the bottom panel of Figure~\ref{fig:threshsw1}.  Any signals in the ground-based data are convolved with this complex aliasing structure in the FT, making it difficult to select the correct pulsation frequencies. The x-axes of both panels of Figure~\ref{fig:threshsw1} have the same scale, emphasizing the relative imprecision of ground-based signal detections.  We aim to use the precise \Ktwo\ data to guide our selection of the correct alias peaks; in doing so, we determine the number of Nyquist bounces of the \Ktwo\ signals and recover accurate pulsation frequencies at \Ktwo\ precision.  When the \Ktwo\ data do not assist our peak selection, we risk choosing the wrong aliases and adopting frequencies that are off by of-order 11.6\,$\mu$Hz, as discussed in Section~\ref{sec:sw}.

\begin{deluxetable}{l r r r}
\tablecolumns{4}
\tablecaption{Significant \Ktwo\ Aliases for EPIC\,210377280\label{tab:k21}}
\tablehead{
\colhead{} & \colhead{Frequency} & \colhead{Period} & \colhead{Amplitude}\\
\colhead{} & \colhead{($\mu$Hz)} & \colhead{(s)} & \colhead{(\%)}}
\startdata
{\color{one}\ding{117}} F$_1$ &	68.569(13) & 14584(3) & 0.17(3) \\
{\color{two}\ding{117}} F$_2$ &	118.297(11) & 8453.3(8) & 0.20(3) \\
{\color{three}\ding{117}} F$_3$ & 122.630(9) & 8154.6(6) & 0.24(3) \\
{\color{four}\ding{117}} F$_4$ &	236.362(13) & 4230.8(2) & 0.17(3)

\enddata
\end{deluxetable}

\begin{figure*}[t]
  \centering
  \includegraphics[width=1.99\columnwidth]{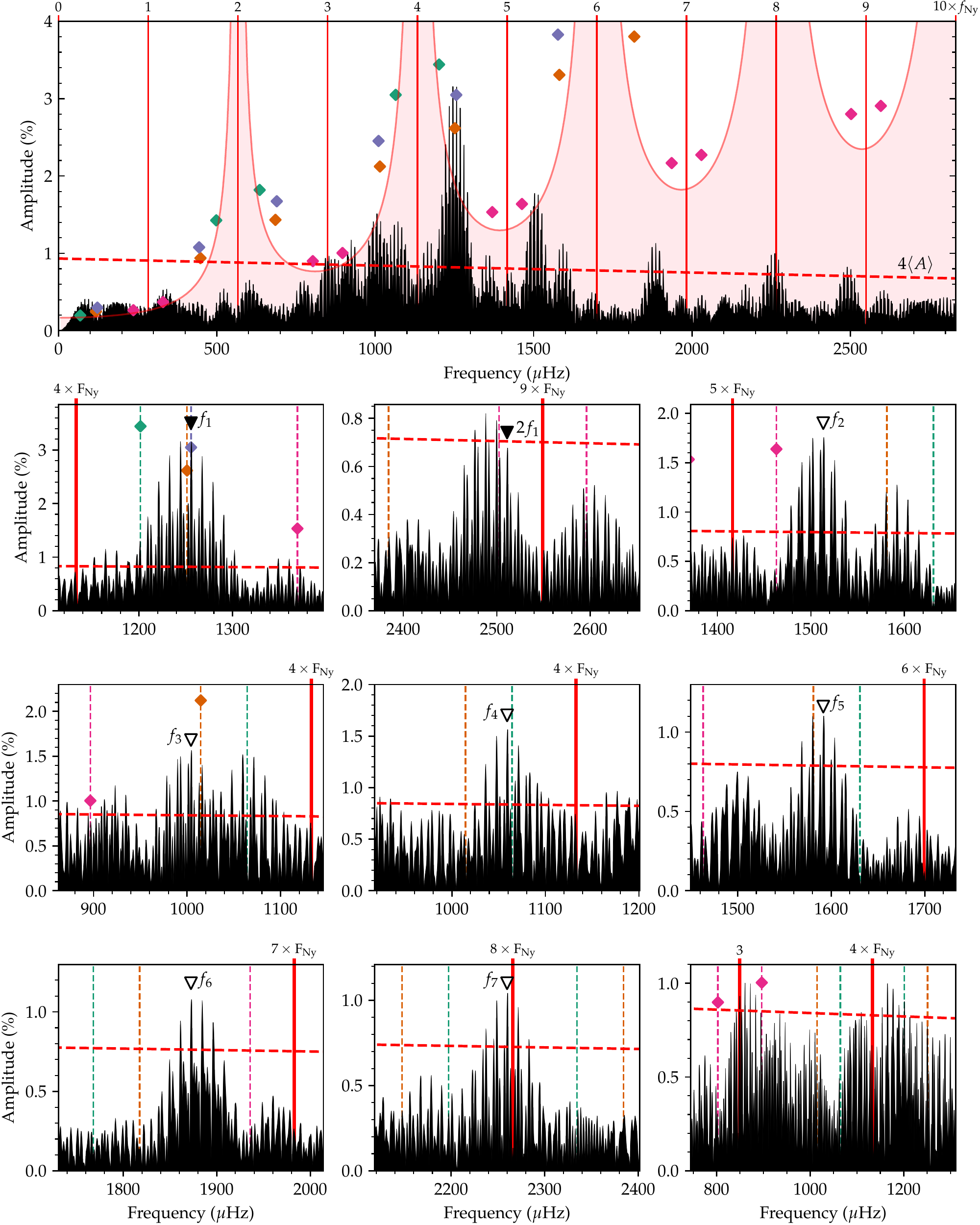}
  \caption{{\sc Top:} Fourier transform of the McDonald observations of EPIC\,210377280 with the \Ktwo\ sensitivity function overlaid (\Ktwo\ observations are not sensitive to the shaded region to a FAP of 0.1\%). Solid vertical lines mark integer multiples of the \Ktwo\ Nyquist frequency.  The possible intrinsic frequencies and expected amplitudes (corrected for phase smearing and bandpass differences) corresponding to the measured aliases in the \Ktwo\ FT are indicated with diamond markers (color coded to match the top panel of Figure~\ref{fig:threshsw1}). {\sc Bottom:} Prewhitening sequence for EPIC\,210377280 (progresses left to right, then down). Black triangles mark alias frequency selections supported by \Ktwo\ observations; unfilled triangles point to peaks selected from ground-based data alone. The red dashed line in all panels shows the $4\langle A\rangle$ significance threshold for the ground-based data.  The last panel highlights the residual power in our fully prewhitened light curve.}
  \label{fig:comp1}
\end{figure*}

\begin{figure*}[t]
  \centering
  \includegraphics[width=2\columnwidth]{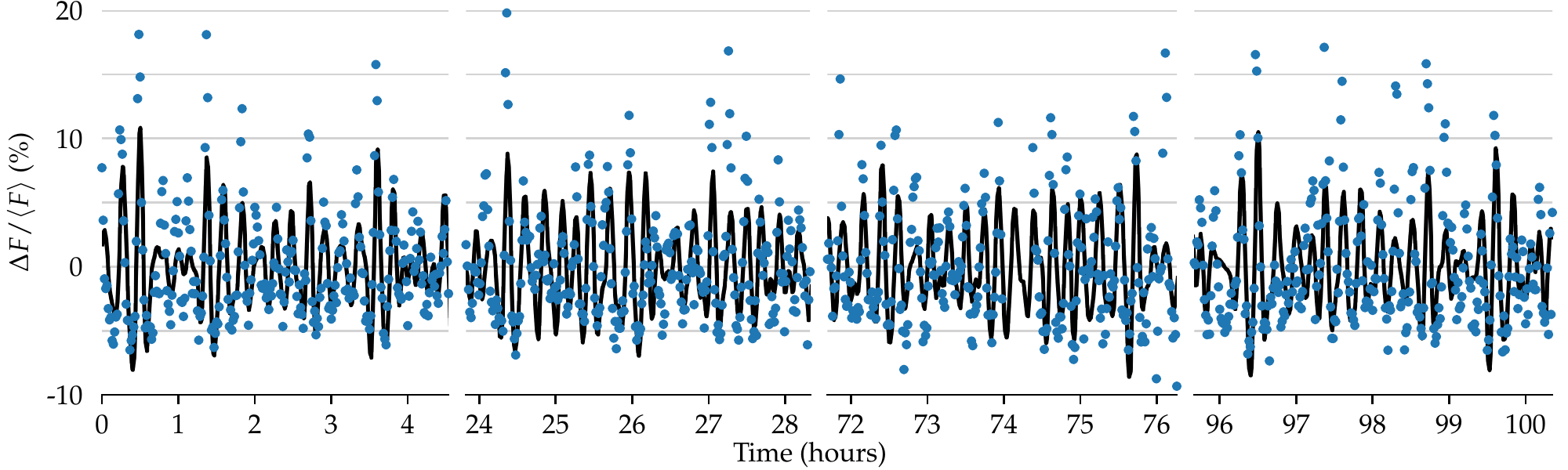}
  \caption{The frequency solution from Table~\ref{tab:mcdf1} overlaid on the ground-based observations of EPIC\,210377280 from McDonald Observatory.}
  \label{fig:lc1}
\end{figure*}

The FT of the ground-based observations of EPIC 210377280 in the full range of pulsational power is presented in the top panel of Figure~\ref{fig:comp1}. The pink shaded region corresponds to the 0.1\% FAP level for the \Ktwo\ data (dashed line in Figure~\ref{fig:threshsw1}), corrected for the effects of phase smearing (Section~\ref{sec:ps}) and passband differences (Section~\ref{sec:pd}). This essentially represents the sensitivity of the long-cadence \Ktwo\ observations to the pulsation signatures measured from the ground; if these light curves were obtained simultaneously, we would expect the signals above the pink shaded region in Figure~\ref{fig:comp1} to rise above the 0.1\% FAP threshold in Figure~\ref{fig:threshsw1}.  Integer multiples of the Nyquist frequency are marked in the figure with vertical red lines, and the $4\langle A\rangle$ significance threshold for the ground-based data is displayed as the dashed red line.  Notably, the \Ktwo\ observations would not be sensitive to the majority of significant pulsations that we detect from the ground.

Figure~\ref{fig:comp1} also indicates the possible intrinsic frequencies and expected amplitudes corresponding to the significant signals measured in the \Ktwo\ data with diamonds that are color-coded to match Figure~\ref{fig:threshsw1} and Table~\ref{tab:k21}.

After adopting each new frequency from the ground-based data, we use a least-squares optimization in \mbox{\sc Period04} to refine the overall solution.  We then subtract (prewhiten) the best fit and search for additional significant signals in the FT of the residuals.  The sequence of FTs in the smaller panels of Figure~\ref{fig:comp1} demonstrates our process of frequency selection. Those signals marked with solid black triangles were informed by the precision \Ktwo\ data, while signals marked with white triangles were selected as the highest-amplitude peaks in the ground-based data and may be the incorrect aliases.

The modes are characterized in Table~\ref{tab:mcdf1} in order of adoption, and this best-fit model is plotted over the McDonald light curve in Figure~\ref{fig:lc1}.  We refer to frequencies detected in the \Ktwo\ data with a capital F$_n$ (as in Table~\ref{tab:k21}), and our final frequencies as $f_n$ (Table~\ref{tab:mcdf1}). For frequencies matched to a \Ktwo\ signal, we refine the frequency value and uncertainty by doing a final least-squares fit of the correct alias to the \Ktwo\ data; otherwise, the values come from least-squares fits to the McDonald light curve.  Formal \emph{intrinsic} uncertainties are quoted following \citet{Montgomery1999}.

\begin{deluxetable}{l r r r}
\tablecolumns{4}
\tablecaption{Frequency Solution for EPIC\,210377280\label{tab:mcdf1}}
\tablehead{
\colhead{Mode} & \colhead{Frequency} & \colhead{Period}  & \colhead{Amplitude\tablenotemark{a}}\\
\colhead{} & \colhead{($\mu$Hz)} & \colhead{(s)} & \colhead{(\%)}}
\startdata
$f_1$\tablenotemark{b} & 1255.581(9) & 796.444(6) & 3.33(14) \\
2$f_1$\tablenotemark{b,c} & 2511.162(19) & 398.222(3) & 0.64(14) \\
$f_2$\tablenotemark{d} &	1513.14(3) & 660.88(6) & 1.66(14) \\
$f_3$\tablenotemark{d} & 1004.17(13) & 995.85(13) &  1.62(14) \\
$f_4$\tablenotemark{d} &	1059.07(13) & 944.22(12) & 1.57(14) \\
$f_5$\tablenotemark{d} & 1591.54(19)  & 628.32(7) & 1.11(14) \\
$f_6$\tablenotemark{d} & 1872.49(19)  & 534.05(5) & 1.09(14) \\
$f_7$\tablenotemark{d} & 2259.8(2) & 442.51(4) & 1.05(14)
\enddata
\tablenotetext{a}{Amplitude based on fit to ground-based data only.}
\tablenotetext{b}{Alias matched to \Ktwo\ signal.}
\tablenotetext{c}{Harmonic of previously found signal.}
\tablenotetext{d}{Detected in ground-based data alone and may be incorrect alias of spectral window (see text).}
\end{deluxetable}

The first small panel of Figure~\ref{fig:comp1} depicts the first significant frequency that we adopt. This is a clear demonstration of the combined strength of these two data sets: one of the highest aliases (but not \emph{the} highest; see discussion in Section~\ref{sec:dc}) of the largest signal from the ground-based photometry precisely matches the four-Nyquist-bounce candidate intrinsic signal underlying F$_3$ in both amplitude and frequency.  We fit and prewhiten this signal from the light curve to search for additional signals in the residuals.  We also identify the second harmonic of this mode, which is displayed in the second small panel of Figure~\ref{fig:comp1}; the exact 2:1 ratio of harmonics informs the selection of the correct peak corresponding to $2f_1$, which is the sixth highest-amplitude alias of that signal.  Since F$_3$ has been positively matched to a ground-based peak, we exclude other possible underlying solutions for this mode in later panels.  Our final frequency constraint on $f_1$ in Table~\ref{tab:mcdf1} has a total uncertainty of only 0.009\,$\mu$Hz.

The other frequencies that we adopt do not unambiguously resolve \Ktwo\ aliases.  For instance, F$_2$ has candidate frequencies close to aliases of both $f_3$ and $f_5$. We simply adopt the highest alias peaks associated with each ground-based signal in Table~\ref{tab:mcdf1}.  At the amplitudes observed in the McDonald data, many of these would not be detectable at long cadence by \Ktwo\ due to amplitude suppression from phase smearing (Section~\ref{sec:ps}). A notable example is $f_7$, which appears very close to $8\times\mathrm{F}_\mathrm{Ny}$.  An exception to this is $f_2$, which was most likely not excited to such high amplitude throughout the \Ktwo\ observations, as discussed in Sections~\ref{sec:coherence} and \ref{sec:dc}. 

The bottom-right panel of Figure~\ref{fig:comp1} displays part of the FT of the final residuals relative to our significance threshold.  While a few remaining peaks exceed our adopted significance criterion, alias selection is non-trivial and not attempted.

\subsection{EPIC\,220274129}
\label{sec:star2}

\begin{figure}[t]
  \centering
   \includegraphics[width=0.9\columnwidth]{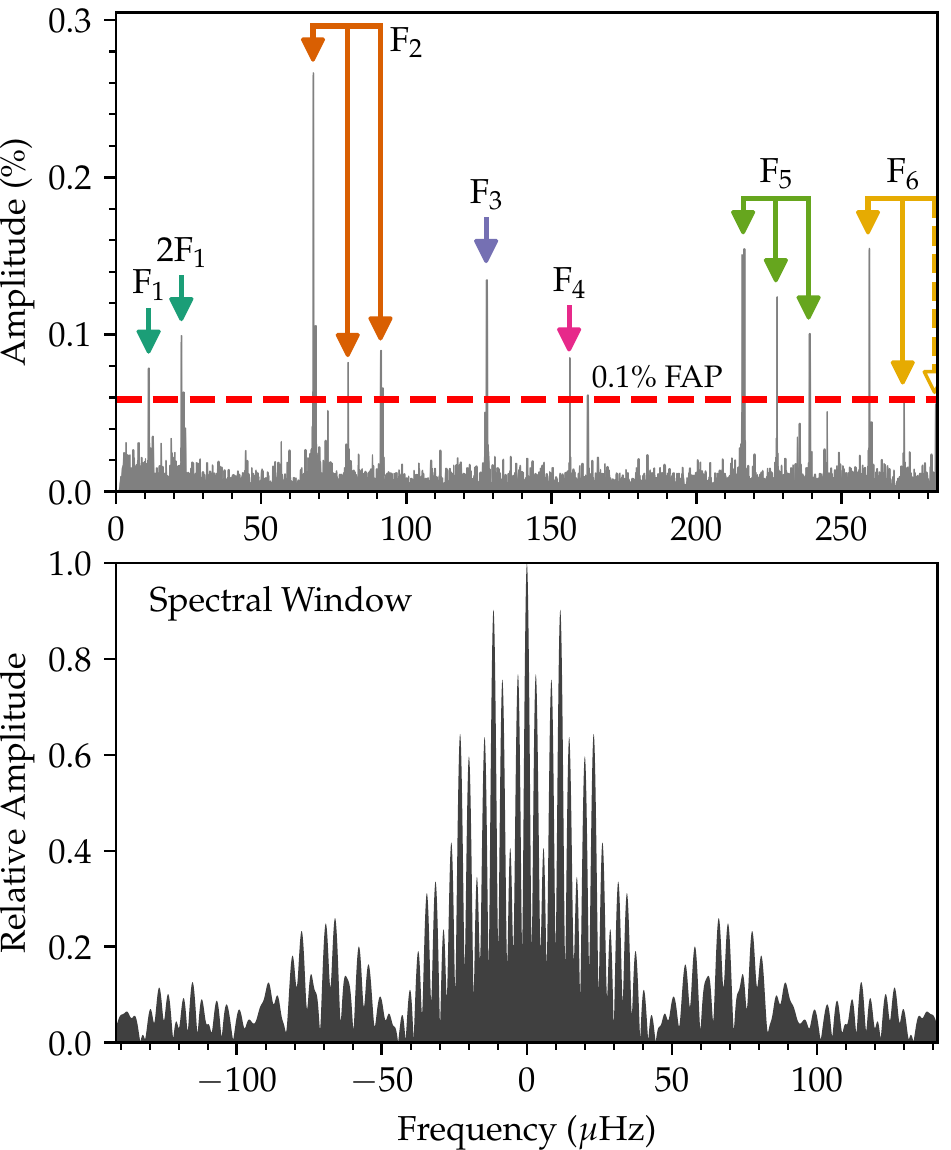}
  \caption{{\sc Top:} Fourier transform of the \Ktwo\ observations of EPIC\,220274129 out to the Nyquist frequency with a 0.1\% false alarm probability (FAP) significance threshold (see text). We identify rotationally split mode triplets with connected arrows (see text). {\sc Bottom:} The spectral window from three nights of McDonald observations of EPIC\,220274129 over a five night span in Oct/Nov 2016. The highest alias is located 11.5\,$\mu$Hz away from the central peak. The x-axes of both panels have the same scale.}
  \label{fig:threshsw2}
\end{figure}

EPIC\,220274129 is 1.7 magnitudes brighter than EPIC 210377280 and exhibits a greater number of signals that exceed the 0.1\% FAP threshold (0.059\% amplitude), as displayed in the top panel of Figure~\ref{fig:threshsw2}.  These are characterized in Table~\ref{tab:k22}.  We are able to exploit the physics of stellar oscillations to recognize relationships between many of the \Ktwo\ signals, greatly simplifying the task of matching them to the ground-based data.

Rotation can break the azimuthal degeneracy of spherical harmonic pulsation patterns, splitting a mode of spherical degree $\ell$ into $2\ell+1$ multiplet components of integer azimuthal order number $-\ell\le m\le\ell$.  For the nonradial gravity mode pulsations of white dwarfs, modes of different $m$ show frequency spacings of
\begin{equation}\label{eq:rot}
\delta\nu_{k\ell m}=(m/P_\mathrm{rot})\times(1-C_{k\ell}),
\end{equation}
where $P_\mathrm{rot}$ is the rotation rate (assumed solid body) and $C_{k\ell}$ is the Ledoux constant that describes the effect of the Coriolis force on a particular mode \citep{Ledoux1951}.  We use the observational convention that $m=+1$ is the higher frequency component.  In the asymptotic limit of high radial order, $k$, $C_{k\ell} \approx 1/\ell(\ell+1)$, but generally $C_{k\ell} \le 1/\ell(\ell+1)$.

We identify among the \Ktwo\ signals three sets of likely rotationally split $\ell=1$ triplets that share a similar frequency spacing with a mean of 11.56\,$\mu$Hz.  These are marked with connected arrows in Figure~\ref{fig:threshsw2}. The highest detected frequency is close enough to the Nyquist that it may have reflected one more or fewer time than the other components of the F$_6$ triplet (dashed arrow in Figure~\ref{fig:threshsw2}), so we exclude it from our calculation of the mean frequency splitting.  Following the approach of \citet{Hermes2017b}, we assume a typical $C_{k\ell} = 0.47$ and a 10\% systematic uncertainty to asteroseismically determine the rotation period of EPIC\,220274129 to be $12.7\pm1.3$\,hr.  This period is typical of white dwarfs with measured rotation rates \citep{Kawaler2015,Hermes2017b}.

The two signals with the lowest frequencies, F$_1$ and 2F$_1$, are consistent with having a 1:2 ratio.  We interpret that these are not super-Nyquist signals and have accurate periods of $24.658\pm0.017$ and $12.324\pm0.003$\,hr.  We discuss the possible nature of these signals at the end of this section, but otherwise exclude them from our search for pulsations in the ground-based data.

The spectral window of our McDonald observations is displayed in the bottom panel of Figure~\ref{fig:threshsw2}. The near coincidence between the location of the highest alias at 11.5\,$\mu$Hz and the splitting of $\ell=1$ multiplets underscores the insensitivity of single-site ground-based observations for asteroseismic measurements of typical white dwarf rotation rates.

Because we have measured the splittings of the $\ell=1$ triplets, we can determine the intrinsic frequencies of each set to \Ktwo\ precision by resolving the super-Nyquist ambiguity of any one component.  For triplets that we match to the ground-based data, we are also able to identify which signals correspond to the $m=+1$ versus $m=-1$ components.

We exclude from Table~\ref{tab:k22} a few peaks that reach significant amplitudes that are within 1\,$\mu$Hz of the higher-amplitude signals associated with F$_2$ (both $m=\pm1$ components) and F$_5$ (lowest frequency component).  These are likely caused by amplitude modulation during the \Ktwo\ observations distributing the power into a set of closely spaced peaks (Section~\ref{sec:coherence}).  Though it is just below our conservative significance threshold, we include the highest frequency component of F$_6$ because it is found at the expected location given the rotational frequency splitting observed in other modes.

\begin{deluxetable}{l r r r}
\tablecolumns{4}
\tablecaption{Significant \Ktwo\ Aliases for EPIC\,220274129\label{tab:k22}}
\tablehead{
\colhead{} & \colhead{Frequency} & \colhead{Period} & \colhead{Amplitude}\\
\colhead{} & \colhead{($\mu$Hz)} & \colhead{(s)} & \colhead{(\%)}}
\startdata
{\color{one}\ding{117}} F$_1$\tablenotemark{a} & 11.266(7) & 88770(60) & 0.081(7) \\
{\color{one}\ding{117}} 2F$_1$\tablenotemark{a} & 22.540(6) & 44360(12) & 0.101(7) \\
{\color{two}\ding{117}} F$_2$ ($m$=$\mp$1)\tablenotemark{b} & 67.901(2) & 14727.4(5) & 0.266(7) \\
{\color{two}\ding{117}} F$_2$ ($m$=0)\tablenotemark{b} & 79.890(8) & 12517.2(1.2) & 0.078(7) \\
{\color{two}\ding{117}} F$_2$ ($m$=$\pm$1)\tablenotemark{b} & 91.159(7) & 10969.8(8) & 0.089(7) \\
{\color{three}\ding{117}} F$_3$ & 127.589(4) & 7837.8(3) & 0.138(7) \\
{\color{four}\ding{117}} F$_4$ & 156.169(7) & 6403.3(3) & 0.082(7) \\
{\color{five}\ding{117}} F$_5$ ($m$=$\mp$1)\tablenotemark{b} & 216.156(4) & 4626.29(8) & 0.156(7) \\
{\color{five}\ding{117}} F$_5$ ($m$=0)\tablenotemark{b} & 227.427(5) & 4397.02(9) & 0.126(7) \\
{\color{five}\ding{117}} F$_5$ ($m$=$\pm$1)\tablenotemark{b} & 238.701(6) & 4189.34(10) & 0.102(7) \\
{\color{six}\ding{117}} F$_6$ ($m$=$\mp$1)\tablenotemark{b} & 259.161(4) & 3858.60(6) & 0.156(7) \\
{\color{six}\ding{117}} F$_6$ ($m$=0)\tablenotemark{b} & 271.145(10) & 3688.07(13) & 0.061(7) \\
{\color{six}\ding{117}} F$_6$ ($m$=$\pm$1)\tablenotemark{b,c} & 282.130(11) & 3544.47(14) & 0.055(7) 
\enddata
\tablenotetext{a}{F$_1$ is a subharmonic of the rotation rate, 2F$_1$.}
\tablenotetext{b}{Component of a rotationally split $\ell$=1 triplet.}
\tablenotetext{c}{This signal may have reflected one more or fewer time than the other components of this triplet (see text).}
\end{deluxetable}

\begin{figure*}[t]
  \includegraphics[width=1.99\columnwidth]{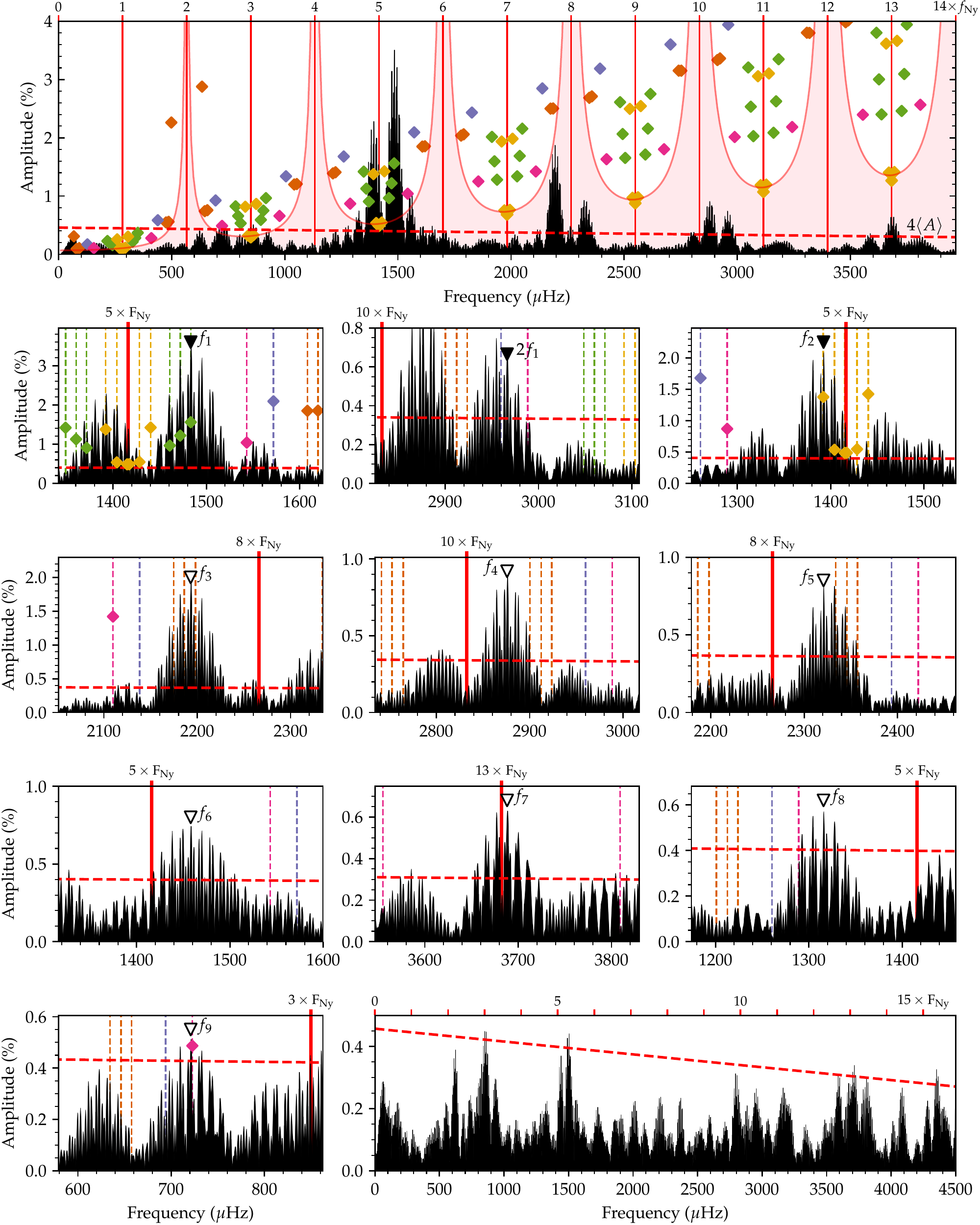}
  \caption{Same as Figure~\ref{fig:comp1}, except for EPIC\,220274129. Mode-by-mode frequency adoption and prewhitening progresses in the sequence of smaller panels from left to right, then top to bottom. The vertical dashed lines and diamonds indicate candidate frequencies and amplitudes underlying the \Ktwo\ signals, color-coded to match the triangles in Figure~\ref{fig:threshsw2}. Signals marked with black triangles were informed by \Ktwo\ data, while unfilled triangles mark signals selected solely from the ground-based observations. The last panel highlights residual power in the fully prewhitened light curve.}
   \label{fig:comp2}
\end{figure*}

\begin{figure*}[t]
  \centering
  \includegraphics[width=2\columnwidth]{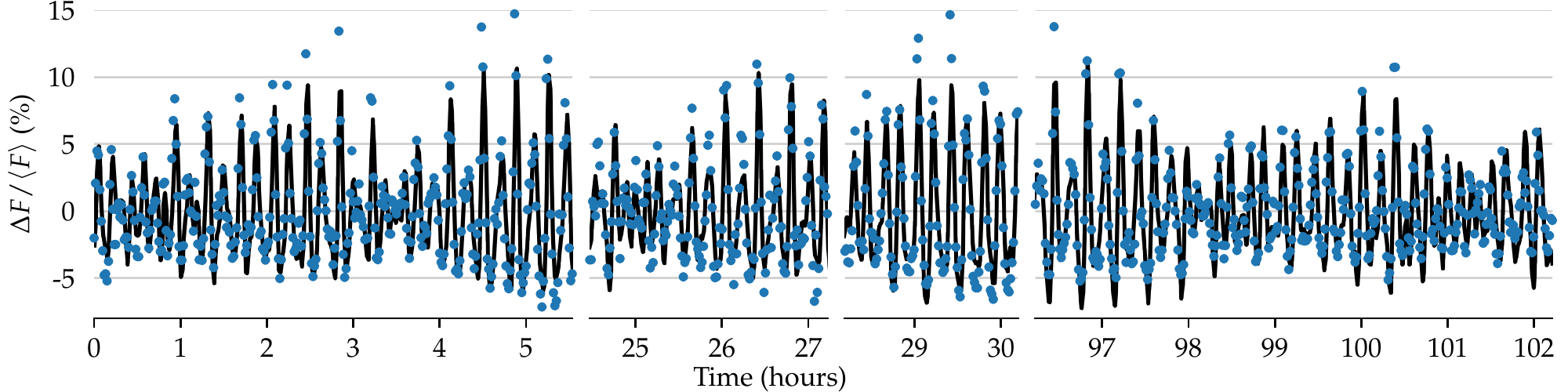}
  \caption{The frequency solution overlaid on the ground-based observations of EPIC\,220274129 from McDonald Observatory.}
  \label{fig:lc2}
\end{figure*}

The top panel of Figure~\ref{fig:comp2} depicts our comparison of possible frequencies underlying the \Ktwo\ pulsation signals to the FT of the McDonald data in the full range of significant pulsational power.  Like Figure~\ref{fig:comp1} for the previous object, the pink shaded region highlights the sensitivity of the \Ktwo\ data, and the diamonds are color-coded to match Figure~\ref{fig:threshsw2} and Table~\ref{tab:k22}.

We adopt and prewhiten pulsation signals from the ground-based data in each of the smaller panels.  First, we recognize that the largest signal from the ground ($f_1$) matches a candidate frequency for the highest-amplitude component of the F$_5$ triplet.  Resolving that these signals were aliased off the Nyquist five times in the \Ktwo\ data, we can refine the frequencies of all three components by fitting the correct signals to the full \Ktwo\ light curve.  These precise frequencies are provided in Table~\ref{tab:mcdf2}.  Because there is a ground-based aliasing degeneracy that matches the $\ell=1$ rotational splitting, we can only prewhiten the single dominant component from the ground based data.  We report a ground-based amplitude for this signal in Table~\ref{tab:mcdf2}, but this may not be physical due to beating against aliases of these other modes.  As with EPIC\,210377280, we also identify the second harmonic of the dominant signal from an exact 2:1 frequency ratio, which we display in the second small panel of Figure~\ref{fig:comp2}.

The third small panel of Figure~\ref{fig:comp2} indicates agreement between the next strongest ground-based signal ($f_2$) and the triplets of the F$_6$ detection.  We use the \Ktwo\ data to refine these fits in Table~\ref{tab:mcdf2}. Because of the proximity of the one of these signals to the \Ktwo\ Nyquist, we must still admit two possible solutions for the $m=+1$ mode, but with the central component clearly identified, this should not affect follow-up asteroseismic fitting.

\begin{deluxetable}{l r r r}
\tablecolumns{4}
\tablecaption{Frequency Solution for EPIC\,220274129\label{tab:mcdf2}}
\tablehead{
\colhead{Mode} & \colhead{Frequency} & \colhead{Period}  & \colhead{Amplitude\tablenotemark{a}}\\
\colhead{} & \colhead{($\mu$Hz)} & \colhead{(s)} & \colhead{(\%)}}
\startdata
$f_1$ ($m$=$-1$)\tablenotemark{b} & 1460.731(6) & 684.589(3) & n/a  \\
$f_1$ ($m$=0)\tablenotemark{b} & 1472.007(5) & 679.345(2) & n/a \\
$f_1$ ($m$=$+1$)\tablenotemark{b} & 1483.276(4) & 674.1832(17) & 3.62(7)  \\
$2f_1$ ($m$=+1)\tablenotemark{b,c} & 2966.553(8) & 337.092(3) &  0.64(7)  \\
$f_2$ ($m$=$-1$)\tablenotemark{b} & 1392.117(4) & 718.331(2) & 1.92(7)  \\
$f_2$ ($m$=0)\tablenotemark{b} & 1404.099(10) & 712.200(5) & n/a \\
$f_2$ ($m$=$+1$)\tablenotemark{b,d} & 1415.085(11) & 706.671(6) & n/a  \\
$f_2$ ($m$=$+1$)\tablenotemark{b,d} & 1417.303(11) & 705.565(5) & n/a  \\
$f_3$\tablenotemark{e}  &	2192.95(5)& 456.006(11) & 1.85(7)  \\
$f_4$\tablenotemark{e}  &	2875.83(11) & 347.725(13) & 0.90(7)  \\
$f_5$\tablenotemark{e}  & 2320.67(12) & 430.91(2) & 0.81(7)  \\
$f_6$\tablenotemark{e,f}  & 1457.98(10) & 685.88(5) & 0.97(7)  \\
$f_7$\tablenotemark{e}  & 3688.13(16) & 271.140(11) & 0.63(7)  \\
$f_8$\tablenotemark{e}  & 1315.75(16) & 760.02(9) & 0.61(7)  \\
$f_9$\tablenotemark{e}  & 720.89(19) & 1387.2(4) & 0.51(7)
\enddata
\tablenotetext{a}{Amplitude based on fit to ground-based data only, where applicable.}
\tablenotetext{b}{Alias refined by \Ktwo\ signal.}
\tablenotetext{c}{Harmonic of previously found signal.}
\tablenotetext{d}{Proximity of the \Ktwo\ signal to $f_\mathrm{Ny}$ admits two candidate solutions to this triplet component.}
\tablenotetext{e}{Detected in ground-based data alone and may be incorrect alias of spectral window (see text).}
\tablenotetext{f}{This peak is suspected to be redundant with the $f_1$ triplet that could not be completely fit out of the ground-based data.}
\end{deluxetable}

The \Ktwo\ data do not clearly prefer a strong alias of the FT structures around the other signals that we detect from the ground. We prewhiten the highest aliases one-at-a-time through $f_9$.  Though $f_9$ does appear to closely match a candidate solution for F$_4$, the match was not compelling before prewhitening many ground-based signals that likely include multiple incorrect alias selections, so we err on the side of caution by not claiming \Ktwo\ precision for this mode.  We also note that $f_6$ falls within the triplet structure of $f_1$, which we were unable to completely fit out of our McDonald light curve because of the degeneracy between rotational splitting and the daily ground-based aliasing; we therefore suspect that this signal is redundant with the $f_1$ triplet properties already refined from \Ktwo\ data.

As with Figure~\ref{fig:comp1}, the last panel of Figure~\ref{fig:comp2} displays residual power that is formally significant but difficult to select aliases from; we assert that additional pulsations are likely excited near these frequencies, but we decline to characterize them from these data. 

Having been aliased to a low measured frequency, the high-amplitude component of the F$_2$ triplet was likely the dominant signal during the \Ktwo\ observations; this is not the dominant signal in the McDonald data, suggesting a significant change in this mode's pulsational power between epochs of observations.

Our final frequency solution for EPIC\,220274129 is listed in Table~\ref{tab:mcdf2}, and the best fit to the McDonald observations is plotted over the light curve in Figure~\ref{fig:lc2}.

\subsubsection{Nature of sub-Nyquist signals}

The signals F$_1$ and 2F$_1$ were likely detected at their intrinsic frequencies and do not represent independent pulsation frequencies.  These were detectable thanks to the extent and high signal-to-noise of the \Ktwo\ data, and while generally beyond the goals of this paper, are deserving of some discussion.  We consider two plausible physical interpretations for these signals: combination frequencies and rotational modulation signatures.

Combination frequencies appear at the sums and differences of ZZ Ceti pulsation frequencies, presumably due to the nonlinear response of the convection zone to pulsations \citep{Brickhill1992,Wu2001}.  The frequencies of the sub-Nyquist signals are similar to the differences between components of $\ell=1$ rotational triplets.

\begin{figure}[t]
  \centering
   \includegraphics[width=.9\columnwidth]{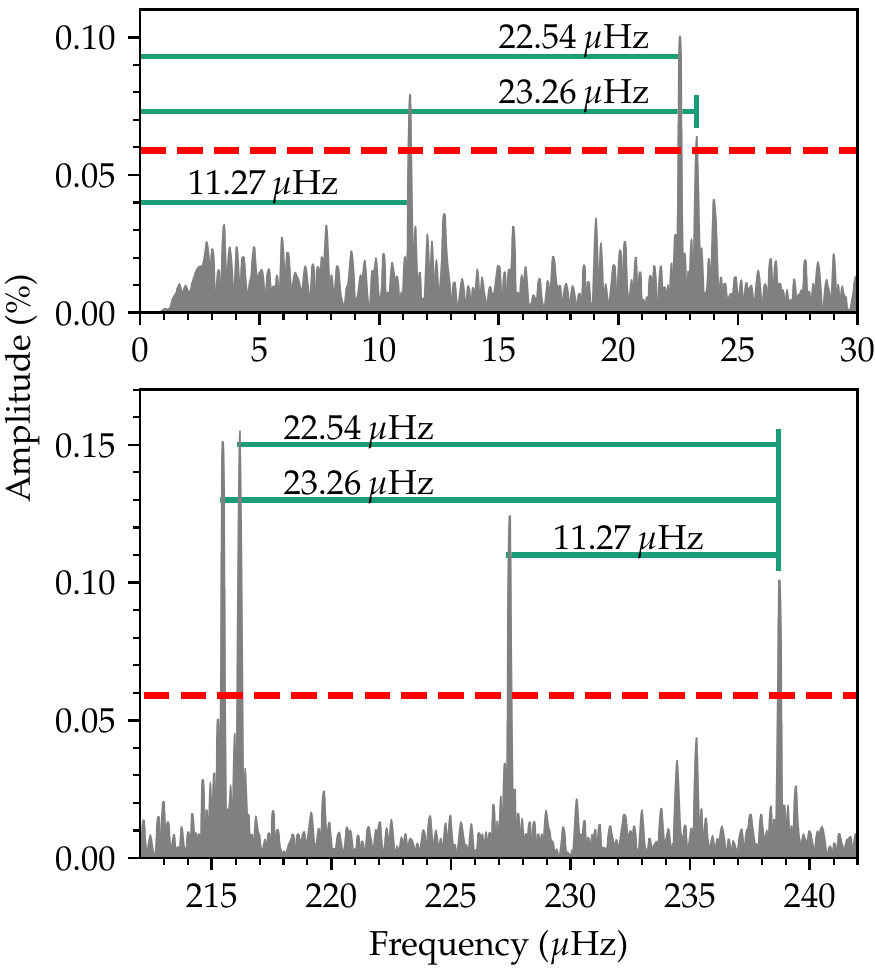}
  \caption{Expanded portions of the Fourier transform of the EPIC\,220274129 \Ktwo\ light curve, showing the {similarities in the structures} of intrinsically sub-Nyquist signals {(F$_1$ and 2F$_1$; top panel)} and the rotationally split multiplets {of the $\ell=1$ F$_5$ pulsations} (bottom panel).}
  \label{fig:subnyq}
\end{figure}

The top panel of Figure~\ref{fig:subnyq} displays these sub-Nyquist signals in greater detail.  The signal near 22.54\,$\mu$Hz actually consists of two closely spaced, significant peaks (with a third compelling but formally insignificant peak nearby).  Individual ZZ Ceti pulsation modes that vary in amplitude during observations appear as clusters of closely spaced peaks in \Kep/\Ktwo\ data (Section~\ref{sec:coherence}). \citet{Hermes2017b} find that long-period pulsations (periods longer than 600\,s) often exhibit these amplitude modulation signatures.  Empirically, modes in this period regime begin to be driven to observable amplitude near the effective temperature of EPIC\,220274129 ($T_\mathrm{eff,3D} = 11810$\,K), which is also approximately where nonlinear combination frequencies reach their largest observed amplitudes in ZZ Cetis \citep{Hermes2017b}. The closely spaced peaks of the lowest-frequency component of F$_5$ (intrinsic period 674.183\,s) displayed in the bottom panel of Figure~\ref{fig:subnyq} are indicative of significant amplitude modulation.  Any combination frequencies involving pulsations that are undergoing intrinsic variations will also vary. For example, the differences between the two peaks of the $m=+1$ component and the $m=-1$ component of F$_5$ mimic the substructure of the 2F$_1$ signal.  While these are compelling candidates for the principle modes of this difference frequency, all $\ell=1$ triplets share similar rotational splittings, including any that were suppressed below our detection threshold by phase smearing. The specific sub-Nyquist structure observed may be the combined signature of multiple difference frequencies in this star.

For $\ell=1$ modes, $C_{kl}\leq 0.5$, and the separation of rotationally split sectoral ($m=\pm1$) components is $\geq P_\mathrm{rot}$ (Equation~\ref{eq:rot}).  While the difference frequency nature of the sub-Nyquist signals is preferred by Occam's razor, we cannot rule out the interpretation {that one of the peaks at 2F$_1$ corresponds to} $1/P_\mathrm{rot}$, manifest by some surface inhomogeneity like starspot modulation. This does not fully explain the observed substructure of the 2F$_1$ signal, as we do not expect photometric rotation to show amplitude modulation on month timescales \citep[the spot modulation of the DA white dwarf BOKS 53856 remained stable throughout 89 days of \Kep\ observations, for instance;][]{Holberg2011}; {therefore, difference frequencies are likely also required to explain the presence of multiple significant peaks near this frequency, as described in the previous paragraph.  If the highest-amplitude component of 2F$_1$ could be confirmed as the rotation rate, this would yield a more precise surface rotation period} of $12.322\pm0.003$\,hr, allowing us to resolve the remaining ambiguity of the $m=+1$ component of the $f_2$ signal (the higher frequency is required for $C_{kl}\leq 0.5$).  This would also enable asteroseismic tests of radial differential rotation models, as well as empirical constraints on values of $C_{k\ell}$ \citep[as proposed for the spot-modulated, pulsating helium-atmosphere (DB) white dwarf PG\,0112+104 by][]{HermesDBV}.  Furthermore, the precise 1:2 ratio of F$_1$ to 2F$_1$ suggests the possible nature of F$_1$ as a subharmonic to the rotation rate. Though such a signal is not expected, F$_1$ could be considered a candidate white-dwarf analog to KIC\,10195926, a roAp star with a rotational subharmonic that \citet{Kurtz2011} suggest represents a torsional (r-mode) oscillation.

\section{Discussion and Conclusions}
\label{sec:dc}

Using a few nights of ground-based photometry, we have successfully resolved the frequency ambiguities underlying a few of the super-Nyquist signals in the long-cadence \Ktwo\ observations of two new ZZ Ceti pulsating white dwarfs, EPIC\,210377280 and EPIC\,220274129.  While these data sets are individually non-ideal for asteroseismic measurements, they complement each other such that we can measure pulsation frequencies to a precision of $\sim0.01\,\mu$Hz. We used a similar approach in \citet{Hermes2017} to confirm the super-Nyquist nature of a rotationally split $\ell =1$ triplet centered on 109.15103\,s in the short-cadence \Ktwo\ light curve of the ZZ Ceti star EPIC\,211914185, but this work is the first example of recovering frequencies beyond 4--5 times the Nyquist.  

By using the \Ktwo\ data to resolve the ground-based aliasing, we have demonstrated the difficulty in selecting the correct peaks from single-site observations with diurnal gaps.  The \Ktwo\ data selected the second most probable alias of $f_1$ from the FT of McDonald data on EPIC\,210377280, rather than the marginally higher-amplitude peak.  For the \Ktwo-informed second harmonics of $f_1$ in both stars, the correct peaks do not have the highest amplitudes locally.  Frequency selection from $<$1\,week of single-site ground-based data alone can be off by many times the daily aliasing of $\approx$11.6\,$\mu$Hz, as many of the frequencies in our solutions certainly are.  This error can be much larger than the offset from assuming the wrong azimuthal order, $m$, which \citet{Metcalfe2003} found to affect but not negate the usefulness of asteroseismic investigations.  Prewhitening by incorrect aliases can also adversely impact the properties of pulsation modes later inferred from the residuals. Any comparison of asteroseismic models to data with gaps should ideally consider the many viable combinations of aliases that can describe the light curves.

Our identification of rotationally split $\ell=1$ triplets in EPIC\,220274129 enabled us to asteroseismically infer the stellar rotation period of $12.7\pm1.3$\,hr.  We also observe sub-Nyquist signals in this star that likely correspond to nonlinear difference frequencies between different components of the same $\ell=1$ multiplets. 

Besides those pulsation signals that we were able to refine to \Ktwo\ precision by positively matching between data sets, we identify $\ge 6$ additional pulsation signatures from the ground-based light curves of each star. These are measured to lower precision and may be inaccurate aliases from the spectral window, but they will still be useful for asteroseismically constraining the interior structures of these rich ZZ Ceti pulsators. The \Ktwo\ data would not have been sensitive to most of these at their observed amplitudes. 

There are also significant signals in the \Ktwo\ data that we were unable to confidently match to the McDonald observations.  As discussed in Section~\ref{sec:coherence}, pulsation amplitudes of ZZ Ceti stars have been observed to change drastically on month timescales, potentially explaining the apparent absence of many of these during ground-based observations. For instance, the highest multiplet component of F$_2$ in EPIC\,220274129 was almost certainly the dominant mode during \Ktwo\ observations, since super-Nyquist signals that are aliased to lower frequencies generally suffer greater amplitude suppression from phase smearing. We also cautiously declined some compelling candidate matches that were not completely unambiguous.

It is possible that the difference between the pseudo-Nyquist frequency and true Nyquist behavior (described in Section~\ref{sec:nyq}) could be exploited to place additional constraints on the intrinsic frequencies of the \Ktwo\ signals as in \citet[][applied to \Kep\ data]{Murphy2013}.  This would provide an important confirmation to our frequency solutions and enable the precise identification of the \Ktwo\ aliases that we did not match in our ground based data.  The addition of these precise frequencies to the solutions for these stars would increase our leverage on asteroseismically constraining their interiors. The sensitivity of this method to \Ktwo\ data, where light curve durations are much shorter than a spacecraft orbit, has not yet been demonstrated (S.~Murphy, private communication).

The spectra and light curves of EPIC\,210377280 and EPIC\,220274129 are available online at \url{http://www.k2wd.org}, alongside the published data for short-cadence ZZ Cetis observed with \Kep\ and \Ktwo\ \citep{Hermes2017b}.

\acknowledgements We thank the anonymous referee for valuable feedback that improved this manuscript. K.J.B., Z.V., M.H.M., and D.E.W.\ acknowledge support from NSF grant AST-1312983. Support for this work was provided by NASA through Hubble Fellowship grant \#HST-HF2-51357.001-A, awarded by the Space Telescope Science Institute, which is operated by the Association of Universities for Research in Astronomy, Incorporated, under NASA contract NAS5-26555. J.T.F.\ and E.D.\ acknowledge support from NSF grant AST-1413001. K.J.B. thanks S.~O.\ Kepler and Simon Murphy for valuable feedback during the preparation of this manuscript.  We thank D.~Koester for sharing his grid of model white dwarf spectra. We acknowledge \Ktwo\ Guest Observer program GO4001 (PI: Kilic) for proposing observations of EPIC\,210377280.  An early draft of this work was included in K.J.B.'s PhD thesis. This paper includes data taken at The McDonald Observatory of The University of Texas at Austin. Additional observations of EPIC\,220274129 on 07 Jul 2016 that confirmed the pulsational nature of this target were made with the assistance of students Zili Shen and Lex Searcy as part of UT-Austin's Freshman Research Initiative White Dwarf Stream's annual summer trip to McDonald Observatory. Based on observations obtained at the Southern Astrophysical Research (SOAR) telescope, which is a joint project of the Minist\'{e}rio da Ci\^{e}ncia, Tecnologia, e Inova\c{c}\~{a}o da Rep\'{u}blica Federativa do Brasil, the U.S. National Optical Astronomy Observatory, the University of North Carolina at Chapel Hill, and Michigan State University. This paper includes data collected by the \Ktwo\ mission. Funding for the \Ktwo\ mission is provided by the NASA Science Mission directorate. Some of the data presented in this paper were obtained from the Mikulski Archive for Space Telescopes (MAST). STScI is operated by the Association of Universities for Research in Astronomy, Inc., under NASA contract NAS5-26555. Support for MAST for non-HST data is provided by the NASA Office of Space Science via grant NNX09AF08G and by other grants and contracts.

\end{document}